\documentclass[amsmath,showkeys,superscriptaddress,onecolumn,11pt]{revtex4}
\usepackage{float}
\usepackage{titlesec}
\usepackage{amsmath}
\usepackage{amsmath}
\usepackage{amsmath}

\DeclareMathOperator{\arctanh}{arctanh}

\DeclareMathOperator{\arccoth}{arccoth}
\usepackage{amssymb}
\usepackage{graphicx}
\usepackage[colorlinks=true,linkcolor=blue,citecolor=blue,urlcolor=blue,breaklinks]{hyperref}
\begin{document}
\title{Effects of classical fluctuating environments on decoherence and bipartite quantum correlations dynamics}
\author{Atta Ur Rahman}
\email{Attazaib5711@gmail.com}
\author{Muhammad Noman}
\author{Muhammad Javed}
\author{Arif Ullah}
\address{Quantum Optics and Quantum Information Research Group, Department of Physics, University of Malakand, Chakdara Dir, Pakistan}
\author{Ming-Xing Luo}
\address{The school of information science and technology, Southwest Jiaotong University, Chengdu 610031, China}
\begin{abstract}
We address the time evolution of the quantum correlations ($QCs$) such as entanglement, purity, and coherence for a model of two non-interacting qubits initially prepared as a maximally entangled bipartite state. We contrast the comparative potential of the classical fields to preserve these $QCs$ in the noisy and noiseless realms. We also disclose the characteristic dynamical behavior of the $QCs$ of the two-qubit state under the static noisy effects originating from the common and different configuration models. We show that there is a direct connection between the fluctuations allowed by an environment and the $QCs$ preservation. Due to the static noisy dephasing effects, the $QCs$ are suppressed, resulting in the separability of the two-qubit entangled state after a finite duration. Here, the $QCs$ decay effects are found much smaller in the common configuration model than that of the opponent. Furthermore, this protection of the $QCs$ under static noise for large intervals is entirely attributable to the existence of the entanglement sudden death and birth phenomenon. Most importantly, we found the bipartite $QCs$ less fragile than the tripartite ones in comparison under the static noise. In the case of the measures, the concurrence is found to be sharper for showing the entanglement sudden death and birth revivals in comparison to the purity and decoherence.
\end{abstract}
\maketitle
\section{INTRODUCTION}
Composite quantum systems have been found more resourceful for quantum information processing in recent years as a result of the non-classical association between its subsystems \cite{Dillenschneider, Costa, Khalid}. Among such non-classical correlations, quantum entanglement, coherence, and purity are one of the attested phenomenon \cite{Horodecki, Bengtsson}. They are considered as one of the important resources in quantum information sciences and are used to distinguish between classical and non-classical correlations realms and have been investigated extensively in the past and strenuous efforts have been made to detect and measure them \cite{Maziero, Walborn, Steffen, Gray, Dicarlo}. Moreover, different techniques have been developed and deployed to preserve these $QCs$ for different potential quantum mechanical protocols. Quantum computing \cite{Briegel}, quantum dense coding \cite{Hao}, quantum teleportation \cite{Venuti}, Global transmissions \cite{Boone}, quantum games \cite{Khan} and quantum thermodynamical engines \cite{Zhang} are few interesting examples to cite in this case. In this respect, dynamics of the quantum systems is one of the successful mechanisms and must be studied with extensive care for the realistic implementation of these applications.\\
Besides, the processing of quantum information by the use of quantum systems is not an independent technique and depends heavily on a large-scale system known as the environment \cite{Liu}. It is possible to model the concept of interaction between a quantum system and the environment into a classical or quantum interaction realm \cite{Bordone, Paz-Silva}. Here, since it offers more degrees of freedom than its counterpart, we would prefer the classical interpretation of the system-environment coupling. The functional deployment of quantum systems dynamics in various environments has shown that it has fatal effects on the preservation of the $QCs$ \cite{Franco-1, Franco-2, Yu}. They are degraded in some situations, while in some worst cases they become disappear. Besides this, multiple features of these environments such as degrading capacity, memory properties, revivals of the entanglement, and short and long-lived correlations conditions have also been investigated \cite{Maziero, Liu, Wang, Horodecki}. Besides, different types of noises arising from the coupling of quantum systems with classical environments were also remained active part of the previous studies \cite{Devetak, Javed, Shamirzaie, Arthur, Lionel}.\\
In the present paper, we investigate the dynamics of bipartite $\mathit{QCs}$ which includes entanglement, purity, and coherence. In this domain, the dynamics characterized by the coherence decay will be measured in terms of the Von Neumann entropy \cite{Boes}. By using concurrence and purity measures, we will show the qualitative behavior of the preserved entanglement and purity for the two-qubit entangled state \cite{Islam-1, Uhlmann}. Employing different types of measures will be effective to provide valid results and observe the dynamics of different kinds of bipartite $QCs$ simultaneously. These $\mathit{QCs}$ are further subjected to static noise stemming from the coupling of the classical fluctuating fields with two non-interacting entangled qubits. This noise exists primarily because of the lattice disorder and has shown fatal effects in various quantum applications, such as quantum walks and static random-access memory technologies \cite{Benedetti-1, Calhoun, Birla, ArthurT, Rahman, Kropf}. Hereafter, we also intend to distinguish the classical noiseless and noisy fields. In the case of the classical fields, fluctuations are considered to be the main characteristic properties of the time evolution of quantum systems. In this case, we will intend to investigate any direct connection between these fluctuations and the preservation of the $QCs$. In turn, it will predict the capacity of a classical environment to preserve $QCs$. Besides this, we will evaluate the functional role of each static noisy parameter and will be presented in an outlook. This characterization would make it easy to prevent degradation of the $\mathit{QCs}$ by setting the parameters optimally. it will provide means for the practical implementation of quantum information processing with enhanced dynamics and preservation of the $\mathit{QCs}$. The qualitative behavior of the dynamics of the $QCs$ will be studied in both strong and weak coupling regimes of the noise. Moreover, the coupling of the bipartite system will be investigated in two different models. In the first model, both the qubits will be coupled to a single environment and are named as common configuration model. In the second configuration, each qubit will be linked to a separate unique environment and is named as a different configuration model.\\
This paper is configured as follows: In Sec.$\ref{measures }$, we introduce estimators to compute bipartite $QCs$ which includes concurrence, purity, and decoherence. In Sec.$\ref{MODEL AND DYNAMICS}$, dynamics of the bipartite state when subject to the noiseless and noisy classical fields will be illustrated. In Sec.$\ref{RESULT AND DISCUSSION}$, we present the analytical results obtained for the physical model. The paper closes with some concluding remarks in Sec.$\ref{CONCLUSION}$. In Sec.$\ref{Appendix}$, we will present the numerical results obtained for the physical model and bipartite quantum correlations.
\section{BIPARTITE QUANTUM CORRELATIONS MEASURES}\label{measures }
Before going into the context of dynamics, we first introduce general estimators for quantifying coherence, entanglement, and purity.
\subsection{Decoherence}
Decoherence ($\mathcal{D}(t)$) defined by the Von Neumann entropy measures the entanglement as well as the coherence decay.  This decay occcurs because of the dissipation of the entanglement due to the involved environmental noise. Here, we can get $\mathcal{D}(t)$ for the time evolved state of a system $\rho(t)$ as \cite{Arthur}:
\begin{equation}
\mathcal{D}(t)=-Tr[\rho_{ab}(t)\ln\rho_{ab}(t)]=-\sum_{i=1}^{4}\nu_i \ln \nu_i, \label{decoherence measure}
\end{equation}
where $\nu_i$ are the eigenvalues of the time evolved density matrix according to the Schmidt decomposition. For the criterion $\mathcal{D}(t)=0$, the state remains entangled while any other value of this measure will indicate the relative decay amount.
\subsection{Purity}
Purity ($\mathcal{P}(t)$) here is included to measure entanaglement and purity for a quantum system. $\mathcal{P}(t)$ for the time evolved density matrix $\rho(t)$ can be computed by \cite{Rispoli}:
\begin{equation}
\mathcal{P}(t)=Tr[\rho_{ab}(t)]^{2}, \label{purity measure}%
\end{equation}
where $ \mathcal{P}(t) $ for n-dimensional state ranges between $\frac{1}{n}$ and $1$. For a pure state $ \mathcal{P}(t) $ is $1$ while for $\frac{1}{n}$, the state becomes completely mixed and separable. 
\subsection{Concurrence}
Concurrence $(\mathcal{C}_e(t))$ is another measure used for the classification of the entanglement and separability for thr bipartite state where it ranges as $1 \geq \mathcal{C}_e (t)\geq 0 $. Here, $\mathcal{C}_e(t)=1$ shows the state to be entangled while for the minimal bound, the state becomes completely separable. In this paper, we will use $\mathcal{C}_e (t)$ to distinguish between bipartite entanglement and the corresponding separability. $\mathcal{C}_e(t)$ for the two-qubit state is given by \cite{Wootters}:
\begin{equation}
\mathcal{C}_e=max \{0,\sqrt{\nu_4}, \sqrt{\nu_3}, \sqrt{\nu_2}, \sqrt{\nu_1} \}, \label{concurrence}
\end{equation}
where $\nu_i$ are the eigenvalues in decreasing order of the time evolved density matrix $\rho(t)$.
\section{MODEL AND DYNAMICS}\label{MODEL AND DYNAMICS}
The present section evalauates the dynamics of two non-interacting but initially maximally entangled qubits under classical fluctuating environments \cite{Benedetti-3}. The classical fields are considered into noiseless and noisy regimes. Moreover, the time evolution of the two independent qubits is studied in two distinct configurations namely as common $(\rho_{ccm})$ and different configurations model $(\rho_{dcm})$. In the first model, both the qubits are coupled to a single environment and in the later case, each qubit is coupled to the independent noisy environment. The dynamics of the two-qubit system in the case of such stochastic fields is governed by the Hamiltonian model defined as \cite{Franco-3}:
\begin{figure}[ht]
\centering
\includegraphics[scale=0.1]{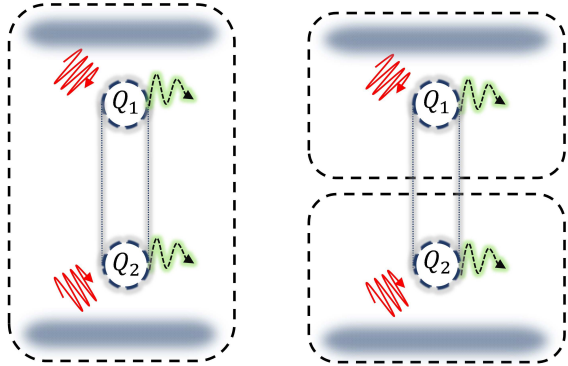}
\caption{Shows the physical model used in this paper for two non-interacting qubits. Here, $ Q_1$ and $Q_2$ represent qubits and the blackish regions show the presence of static noise with square boundaries representing classical environments. The red wavy lines indicate the interaction between the static noisy channels and each qubit. The lines connecting the qubits shows a single composite state composed of two entangled subsystems. The first figure shows the coupling of the two-qubits to a common configurations model (left) and different configuration model (right). The greenish lines show the dynamics of the two-qubits in their corresponding environments.}
\end{figure}
\begin{equation}
\mathcal{H}_{ab}(t)=\mathcal{H}_{a}(t)\otimes I_b+I_{a}\otimes\mathcal{H}_{b}(t),\label{A9}%
\end{equation}
where $\mathcal{H}_{n}(t)$ denotes the corresponding Hamiltonian of the individual qubit and reads as:
\begin{equation}
\mathcal{H}_{n}(t)=\mathcal{E}_n(t) I_n+\lambda\Delta_{n}(t)\sigma_x,\label{Hn}
\end{equation}
where $\mathcal{E}_n$ is the energy associated with the $n$-th qubit and $\lambda$ is the system-environment coupling constant. $\Delta_{n}(t)$ is the stochastic parameter controlling the classical fields, $\sigma_{x}$ is the Pauli matrix and $I $ is the identity operator acting on the subspace of the bipartite state. Hereafter, the time evolved density matrix of the bipartite state is given by \cite{Yu}:
\begin{equation}
\rho_{ab}(t)=U_{ab}(t)\rho_{o}U^{\dag}_{ab}(t), \label{final density matrix}%
\end{equation}
where $\rho_o$ is the initial denisty matrix for the two qubit maximally entangled state and can be written as \cite{Munro,Gonzalez}:
\begin{equation}
\rho_o=r\vert\psi_{ab}^+\rangle \langle\psi_{ab}^+\vert+\frac{(1-r)I_{4\times4}}{4},\label{Initial density matrix}
\end{equation}
here, $r$ is the initial purity of the two-qubit state with $r\in \{0,1\}$ and $ \vert \psi_{ab}^+ \rangle =\frac{1}{\sqrt{2}}({\vert 00 \rangle+\vert 11 \rangle})$ \label{two-qubit}.
Moreover, $U_{ab}$(t) is the time unitary operator and is defined by \cite{Javed}:
\begin{equation}
 U_{ab}(t)=exp \left[-i \int_0^t \mathcal{H}_{ab}(s)ds \right] .\label{U(t)}
\end{equation}
Herefater, we have set $\hbar=1$. Next, the framework for implementing the static noise is pesented. This noise is characterized mainly by the independent noise parameter $ \Delta_m $ having probability distribution given by $ \mathcal{P}(\Delta_x)=\frac{1}{\Delta_m}$. This distribution ranges as $ |\Delta_x-\Delta_o| \leq \frac{\Delta_m}{2} $ and vanishes for any other choices while $ \Delta_o $ represents the mean value of the probability distribution function. Thus, in the case of classical static noise, we get four considerable noise parameters that are $\Delta_m$, $\lambda$, $\Delta_o$ and $t$. The auto-correlation function of the static noise for the noise parameter $ \Delta(f)$ is given by $\langle\delta\Delta(t)\Delta(o)\rangle=\frac{\Delta^{2}_m}{12}$. Next, the impact of the static noise can be found by integrating the final density matrix over the stochastic noise parameter between $\Delta_o-\frac{\Delta_m}{2}$ and $\Delta_o+\frac{\Delta_m}{2}$. By using $Eq.\eqref{final density matrix}$, we get the static noisy effects for $\rho_{ccm}$ as \cite{Lionel}:
\begin{equation}
\rho_{ab}^{ccm}=\int^{\chi_o+\frac{\Delta_m}{2}}_{\chi_o-\frac{\Delta_m}{2}}\frac{\rho_{ab}(t)}{\Delta m}d\Delta a\label{staticcommon}
\end{equation}
where $\Delta a=\Delta b$. For $dcm$, we integrate the $Eq.\eqref{final density matrix}$ as \cite{Lionel}:
\begin{equation}
\rho_{ab}^{dcm}=\int^{\chi_o+\frac{\Delta_m}{2}}_{\chi_o-\frac{\Delta_m}{2}}\int^{\chi_o+\frac{\Delta_m}{2}}_{\chi_o-\frac{\Delta_m}{2}}\frac{\rho_{ab}(t)}{\Delta m^2}d\Delta a d\Delta b.\label{staticmixed}
\end{equation}
\section{RESULTS AND DISCUSSION}\label{RESULT AND DISCUSSION}
In this section, we provide the main results for the dynamics of the two-qubit state when subjected to the classical fields. We have analyzed these classical fields in noiseless and noisy regimes by using the $Eqs.\eqref{final density matrix}$, $\eqref{staticcommon}$ and $\eqref{staticmixed}$. The results for the static noise are further evaluated for the decoherence, purity and concurrence measures from $Eqs.\eqref{decoherence measure}$, $\eqref{purity measure}$ and $\eqref{concurrence}$.
\subsection{Noiseless classical fields}
The dynamics of the two-qubit state, when coupled to noiseless classical fields is presented here. Although, no such fields exist, however, this kind of analysis will be helpful to understand the actual fluctuating character of the classical fields and the effects of each involved parameter. Here, we will focus to relate any direct connection between the existence of the fluctuations and preservation of the $QCs$ for the two-qubit state.
\begin{figure}[ht]
\includegraphics[scale=0.08]{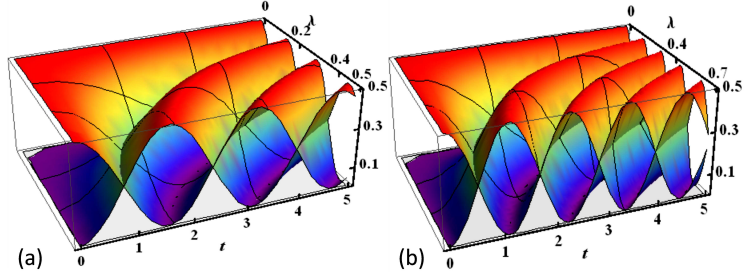}
\caption{Time evolution of the phase of the two-qubit entangled state showing the fluctuation rate when subjected to the classical fluctuating field with noiseless channels for $\lambda =0.5$ (a) and $1$ (b) with $\Delta a=1$ against the evolution parameter $t=5$.}\label{final density matrix without noise}
\end{figure}
$Fig.\ref{final density matrix without noise}$ reports the fluctuation rate for the dynamics of the two-qubit state when subjected to classical fields by using $Eq.\eqref{final density matrix}$. The current results are obtained for $\lambda =0.5$ (a) and $1$ (b) with $\Delta a=1$ against the evolution parameter $t=5$ when subject to $\rho_{ccm}$. In the current case, the superposition of the noise phase over the system phase is not applied and the dynamics is only based on the joint coupling of the phase of the system and stochastic field. It is now obvious that any kind of non-local correlations encoded in a quantum system when linked to such fields will undergo fluctuations due to the random fluctuating properties of this field. These fluctuations are the actual means for the entanglement as well as for other related $QCs$ evolution. Moreover, the fluctuation rates increase for further increase in $\lambda$ as the revival rate is much higher for large values of this parameter relatively. This control of varying the fluctuations in a noiseless classical system may lead to the required optimal characterization of the $QCs$ dynamics. Here, for any values of the parameter, the amplitude of the oscillations remains the same and no damping is seen to be caused by these noiseless environments. Note that we have presented the fluctuations caused in both the real and imaginary parts of the time evolved density matrix to better understand the qualitative behavior of the two-qubit state. Besides this, the interchange of the values between $\lambda$ and $\Delta a$ shows the same results for the $\rho_{ccm}$. Hereafter, the investigations for the fluctuation rate for the coupling of the two qubits with $\rho_{dcm}$ in terms of $\Delta a=\Delta b=1$ with $\lambda=0.5$ and $1$ against the evolution parameter $t$ also shows the same results. Thus, from the current results, one can easily conclude that the detrimental effects can only be due to the noises and not due to such stochastic local fields.
\subsection{Noisy classical fields}
This section includes the dynamics of the two-qubit state when subject to classical fields characterized by the static dephasing noise. Here, we focus on the revival decay rate for the time evolution of the two-qubit state under static noise and will be compared with the results given for noiseless classical fields.\\
\begin{figure}[ht]
\centering
\includegraphics[scale=0.08]{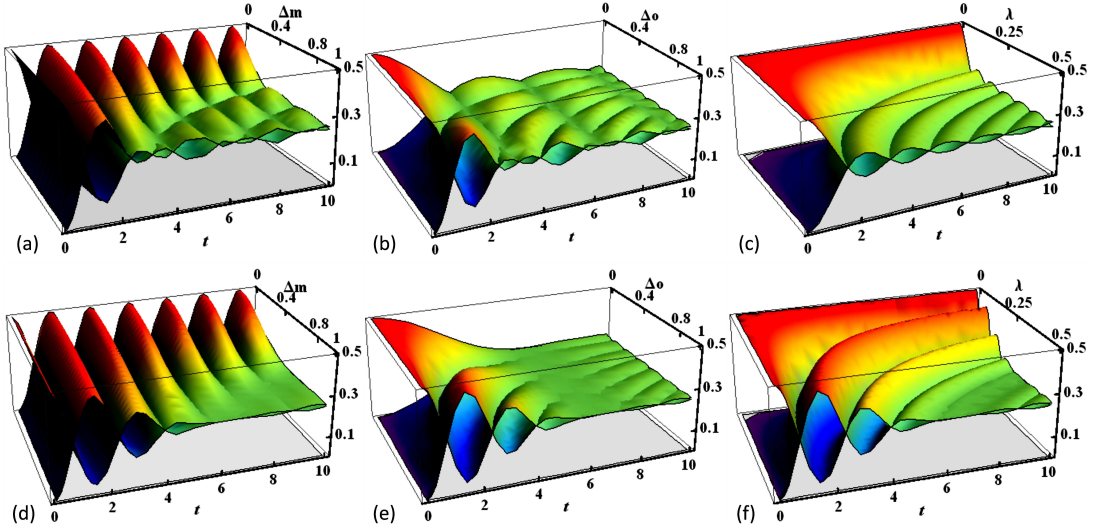}
\caption{Upper panel: Time evolution of the phase of the two-qubit entangled state showing the fluctuation rate when subjected to classical field generating static noise when  $\Delta_m=1$ (first column), $\Delta_o=1$ (second column) and $\lambda=0.5$ (third column) against the evolution parameter $t=10$ when subjected to common configuration model.  Bottom panel: Same as upper panel but for different configuration model.}\label{fianal matrices involving noises}
\end{figure}
In $Fig.\ref{fianal matrices involving noises}$, the qualitative behavior of the fluctuations for the two non-interacting qubits in classical random fields generating static noise is investigated. The current results are obtained by superimposing the noise phase over the system's phase by using $Eqs.\eqref{staticcommon}$ and $\eqref{staticmixed}$ for $\rho_{ccm}$ and $\rho_{dcm}$ respectively. In the current figure, we have evaluated the revival capacity allowed by the stochastic fields for the two-qubit state in terms of the involved parameters. This in turn will show the individual character of each noisy parameter and can be helpful to obtain the optimal fluctuation rate. Here, one can distinguish between the noiseless and noisy classical fields where the time evolution of the fluctuations suffers continuous damping effects due to the static noise. Moreover, the two-qubit state when subjected to $\rho_{ccm}$ and $\rho_{dcm}$ generating static noise exhibits different fluctuating behavior. This implies that the $QCs$ of a quantum system subject to these two environments will show different qualitative dynamics. Besides this, the $\rho_{ccm}$ is found to have a better tendency to preserve fluctuations than $\rho_{dcm}$. This shows correspondence with the previous results analyzing the dynamics of various types of $QCs$ when subject to $\rho_{ccm}$ showing higher fluctuation rate \cite{Liu, Bordone, Arthur, Lionel, Benedetti-3, Rossi, Benedetti-5}. However, the amplitudes of the fluctuations remained higher in $\rho_{dcm}$ than for the $\rho_{ccm}$. Furthermore, we have noticed an increase in the damping rate with the increasing values of the parameters $\Delta m$ and $\lambda$. This implies that the fluctuations can be preserved by turning these two parameters to utmost low values. Besides damping, the parameter $\Delta m$ is witnessed to reduce the amplitude of the fluctuations causing a considerable decrease in the $QCs$. In the case of damping effects, the parameter $\Delta o$ remained non-effective however, this parameter caused the amplitude to increase and has the tendency to control the revival rate for the dynamics of the two-qubit state.
\subsection{Dynamics of the decoherence, purity and concurrence}[ht]
In this section, we give detail description of the time evolution of the decoherence, purity and concurrence by using the $Eq.\eqref{decoherence measure}$, $\eqref{purity measure}$ and $\eqref{concurrence}$. Moreover, the current time evolution of the $QCs$ will be investigated for different values of the noise parameters of the static noise.
\subsubsection{The impact of $\lambda$ over the dynamics of the two qubit state}
\begin{figure}[ht]
\centering
\includegraphics[scale=0.08]{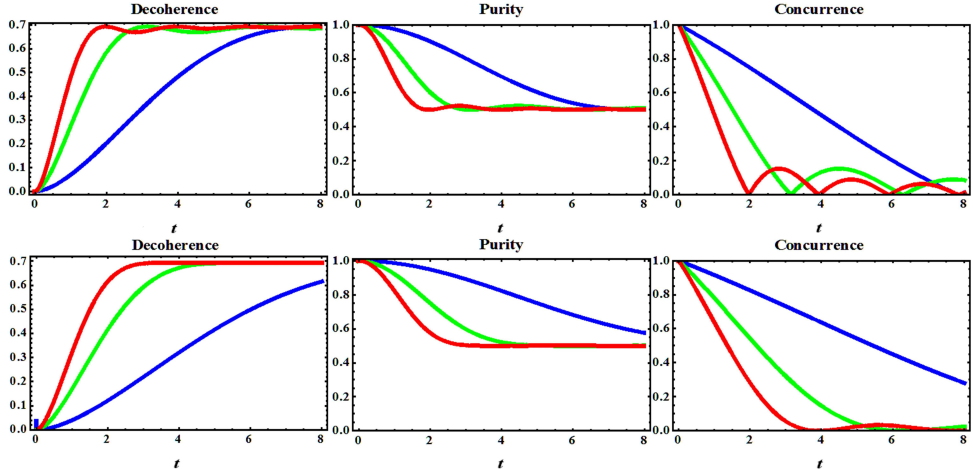}
\caption{Upper panel: Time evolution of the decoherence (first column), purity (second column) and concurrence (third column) for the two qubit entangled state when subjected to common configuration model with $\lambda=0.2$ (blue slopes), $0.5$ (green slopes) and $0.8$ (red slopes) against the evolution parameter $t=8$. Bottom panel: Same as the upper panel but for different configuration model.}\label{lambda discususion}
\end{figure}
$Fig.\ref{lambda discususion}$ evaluates the dynamics of the decoherence, purity, and concurrence for the two-qubit state with different values of $\lambda$ when subject to common (upper panel) and different (bottom panel) configuration models. The static noise has the dominant degrading character of the $QCs$, which is apparent from the current results. Here, the blue-lined slopes shift towards the red end with the increasing order of $\lambda$, inferring that increasing this parameter leads to a greater decay of $QCs$. From the concurrence measurement, the saturation levels reach maximum decay with continuous revivals. Here, the entanglement sudden death ($ESD$) and birth ($ESB$) effects are observed more evident when qubits are coupled to $\rho_{ccm}$. This implies that, for a finite time interval, $QCs$ of the bipartite state faces temporary decay when subjected to $\rho_{ccm}$ and a more permanent decay in the case of $\rho_{dcm}$. Also, the parameter $\lambda$ weakens the initially encoded $QCs$ making the two-qubit state almost separable, particularly in the case of $\rho_{dcm}$. In the case of purity, the two qubits remained entangled for a finite interval but ultimately seems becoming separable. Besides, the initially encoded $QCs$ in the two-qubit state undergo faster decay in $\rho_{ccm}$ due to the static noise compared to $\rho_{dcm}$, however, due to the ESD and ESB revivals, the $QCs$ in $\rho_{ccm}$ were found to be protected for a longer time. On the other hand, the $QCs$ are permanently lost at the first death because of the monotonous decay in addition to the later decay in the $dcm$. Besides this, not only the noise but also the nature of the configuration involved is the reason for affecting the mode of the entanglement decay. This suggested that, due to the static noise, both configurations showed distinct characteristic memory properties. Moreover, the current outlook of the time evolution of the $QCs$ is different enough from the dynamics under different kinds of classical and quantum noises \cite{Aaronson, Mazzola, Karpat, Benedetti-5, Rossi, Solenov}. Apart from this, we found the bipartite $QCs$ acting differently in their qualitative dynamics and preservation as compared to the tripartite $QCs$ under the same noise \cite{Lionel}. In particular, we found the bipartite $QCs$ preserved more than the tripartite ones under the same values of the noise parameters of the static noise in $\rho_{ccm}$ given in \citep{Kenfack-T}. Besides this, the qualitative dynamics is also much different when the static noise is applied in the local mixed noise case for the two-qubit and qutrit-qutrit entangled state \cite{Javed, ArthurT}.
\subsubsection{The impact of $\Delta m$ over the dynamics of the two qubit state}
\begin{figure}[ht]
\centering
\includegraphics[scale=0.08]{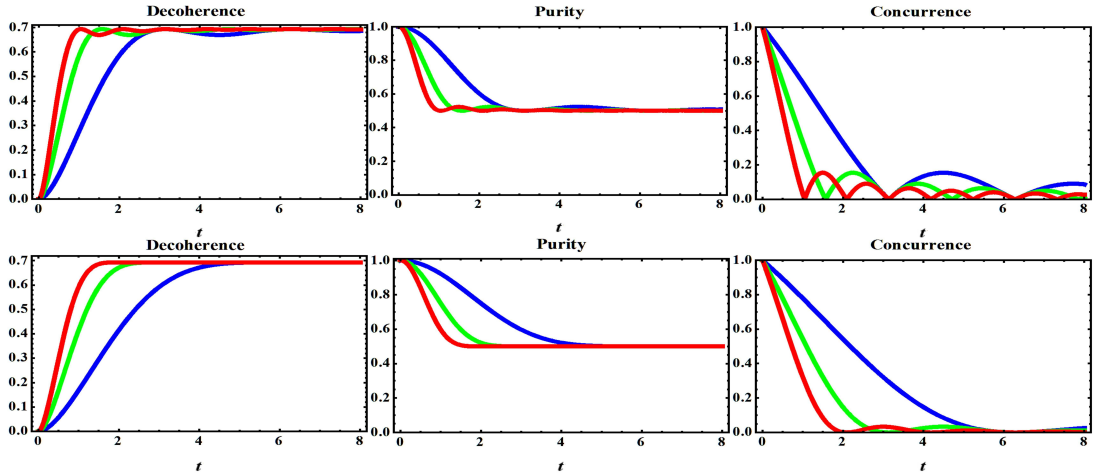}
\caption{Upper panel: Time evolution of the decoherence (first column), purity (second column) and concurrence (third column) for the two qubit entangled state when subjected to common configuration model with $\Delta m=1$ (blue slopes), $2$ (green slopes) and $3$ (red slopes) against the evolution parameter $t=8$. Bottom panel: Same as the upper panel but for different configuration model.}\label{DELTAM}
\end{figure}
$ Fig.\ref{DELTAM} $ explores the dynamics of the $QCs$ for the two-qubit state in terms of different values of $\Delta m$ for common (upper panel) and different (bottom panel) configuration model. The findings demonstrate the presence of the dephasing effects due to static noise for the phase of the two-qubit entangled state. The $QCs$ are degraded in different styles in $\rho_{ccm}$ and $\rho_{dcm}$ and seems to disappear at final notes either with monotonic or with revival character. Here, the blue-lined slopes shift towards the red end for the increasing value of $\Delta m$. Therefore, showing the increase in the dissipation of the entanglement for the increases in the parameter $\Delta m$. Also, obvious $ESD$ and $ESB$ effects are observed for the entanglement dynamics particularly against $\Delta m$. As a fact, one can see that the oscillations are greater and are more evident in the case of red-lined slopes for higher values of $\Delta m$. 
\begin{table}[ht]
\tiny{\begin{tabular}{|c|c|c|}
\hline
Parameter & qualitative behavior &remarks \\
\hline
$\Delta m$ in & degrading character  & tendency to increase $ESD$ and $ESD$ revivals\\
noisy regime &&\\
\hline
$\Delta m$ in & strongly controls  & strongly increases the damping effect\\
noiseless regime & amplitudes &\\
\hline
\hline
$\lambda$ in & degrading character	  & directly increases the noisy effects\\
noisy regime &&\\
\hline
$\lambda$ in& no effect over amplitude	  & directly increases the damping effect\\
noiseless regime & &\\
\hline
\hline
$\Delta o$ in& non-effective & directly increases fluctuations\\
noisy regime  &&\\
\hline
$\Delta o$ in& controls amplitude & directly increases fluctuations\\
noiseless regime &of fluctuations &\\
\hline
\hline
$\tau$ in& degrading character & $ESD$ and $ESB$ are more evident for small $\tau$\\
noisy regime  &&\\
\hline
$\tau$ in& damping character & directly increases the damping effects\\
noiseless regime &  &\\
\hline
\end{tabular}}
\caption{Shows the comparative analysis of the noise parameters involved. The qualitative behaviour of each parameter is taken from all of the above findings. The current details in table $\ref{Table4}$ are derived by comparing $Figs.\ref{final density matrix without noise}$, $\ref{fianal matrices involving noises}$, $\ref{lambda discususion}$ and $\ref{DELTAM}$.}\label{Table4}
\end{table}
Moreover, the first minima of the concurrence and purity correspond to the first maxima of the decoherence showing strict agreement in results. Besides all the degrading effects of static noise, the dynamics of the $QCs$ is also significantly influenced by the type of configuration involved. Here, according to the employed measures, the decay noticed is faster enough in the $\rho_{ccm}$. However, one cannot disregard the fact that $QCs$ remained more preserved due to the revivals in $\rho_ {ccm}$, consequently promising to be more appropriate for the survival of the $QCs$. This is due to the temporary loss that resulted in $\rho_{ccm}$, where the information is again shared with the system, hence showing the back-flow of information to the system again from the environment. In the context of $\rho_ {dcm}$, the decay witnessed is nearly monotonous and no repeated revivals were seen after the first death of the slopes except in the case of concurrence. In an outlook of the dynamics of the $QCs$, the qualitative behavior of both configuration models is rather much different. Here, we found the $\Delta m$ not as effective for the degradation of the bipartite $QCs$ as compared to the tripartite $QCs$ \cite{Kenfack-T}. The bipartite $QCs$ especially in the $\rho_{ccm}$ remained preserved for longer duration showing a better capability than the preservation time of the tripartite ones.\\
\section{CONCLUSION}\label{CONCLUSION}
The dynamics and preservation of bipartite quantum correlations ($QCs$) under the decoherence effects of static noise have been investigated. The dynamics of the $QCs$ is assessed in two distinct schemes: common and different configuration models. In the first model, both qubits are linked to a single fluctuating classical environment, while in the second, each qubit is linked to a separate environment. Besides, we have also compared the classical environments to be in noiseless and noisy regimes. The optimal protection of the initial encoded $QCs$ by characterization of the environment and noise also remained a focus of the present study.\\
We found the dynamics of the two-qubit entangled state in the classical environments completely characterized by strong fluctuations. These fluctuations can be controlled by the optimal fixing of the involved parameters. Besides this, no damping effects have been observed for the joint coupling of the bipartite state and noiseless classical fields. We found that $QCs$ are degraded due to static noise, and in almost all the classical noisy fields, it seems to be lost after a finite time interval. In the case of the parameters, we found that $\lambda$ and $\Delta m$ showed a decoherent nature towards the $QCs$ while $\Delta o$ remained ineffective, especially in the case of noisy regimes. Also, for different values of the parameters, there is a single degree of saturation for each metric indicating consistency in results and the same amount of loss.\\
Furthermore, different configuration model showed a later decline for the $QCs$, but due to the entanglement revivals, $QCs$ remained more preserved in the common configuration model. In comparison, the preservation time observed for the bipartite $QCs$ was longer than it is in the case of tripartite states for the same values of the static noisy parameters \cite{Kenfack-T}. It is also realized important that there exists an intrinsic relation between the fluctuations permitted by an environment and the preservation of the $QCs$ by a quantum system. This means environments having a higher capability to preserve fluctuations turn out to be a good resource for the survival of the $QCs$. In the case of $QCs$ measures, there is an exact correspondence between the maxima and minima of these measures showing strict connections. However, in contrast, by detecting greater entanglement revivals, concurrence turns out to be more useful for investigating more thoroughly the qualitative dynamics of the bipartite quantum systems.\\
From our results, we conclude that the $QCs$ encoded initially can be preserved for a longer duration through producing entanglement sudden death and birth revivals. In our case, to enhance the fluctuation rate and preservation of the $QCs$ with revivals, one must deploy a common configuration model with small values of the parameters.
\section{APPENDIX}\label{Appendix}
This section presents the numerical simulations obtained for the dynamics of the two non-interacting entangled qubits when subjected to local stochastic fields. In this case, the time evolution matrix for $\psi_{ab} $ given in $Eq. \eqref{U(t)}$ reads as:
\begin{equation}
U_{ab}(t)=e^{-2 i t \epsilon } \left[
\begin{array}{cccc}
C_{\Delta a} C_{\Delta b} & -i C_{\Delta a} S_{\Delta b} & -i C_{\Delta b} S_{\Delta a} & -S_{\Delta b} S_{\Delta a} \\
 -i C_{\Delta a} S_{\Delta b} & C_{\Delta a} C_{\Delta b} & -S_{\Delta b} S_{\Delta a} & -i C_{\Delta b} S_{\Delta a} \\
 -i C_{\Delta b} S_{\Delta a} & -S_{\Delta b} S_{\Delta a} & C_{\Delta a} C_{\Delta b} & -i C_{\Delta a} S_{\Delta b} \\
 -S_{\Delta b} S_{\Delta a} & -i C_{\Delta b} S_{\Delta a} & -i C_{\Delta a} S_{\Delta b} & C_{\Delta a} C_{\Delta b} \\
\end{array}
\right].
\end{equation}
Next, by using $Eq. \eqref{final density matrix}$, we get the time evolved density matrix of the given state as:
\begin{equation}
\rho_{ab}(t)=\frac{1}{2} \left[
\begin{array}{cccc}
 \cos ^2[\chi_{AB}] & \frac{1}{2} i \sin [2\chi_{AB}] & \frac{1}{2} i \sin [2\chi_{AB}] & \cos ^2[\chi_{AB}] \\
 -\frac{1}{2} i \sin [2\chi_{AB}] & \sin ^2[\chi_{AB}] & \sin ^2[\chi_{AB}] & -\frac{1}{2} i \sin [2\chi_{AB}] \\
 -\frac{1}{2} i \sin [2\chi_{AB}] & \sin ^2[\chi_{AB}] & \sin ^2[\chi_{AB}] & -\frac{1}{2} i \sin [2\chi_{AB}] \\
 \cos ^2[\chi_{AB}] & \frac{1}{2} i \sin [2\chi_{AB}] & \frac{1}{2} i \sin [2\chi_{AB}] & \cos ^2[\chi_{AB}] \\
\end{array}
\right]\label{Final explicit density matrix}
\end{equation}
where
\begin{align*}
C_{\Delta a}C_{\Delta b}=&\cos(\Delta a \lambda  t) \cos (\Delta b \lambda t),& C_{\Delta a} S_{\Delta b}=&\cos (\Delta a \lambda  t) \sin (\Delta b \lambda  t),\\
C_{\Delta b} S_{\text{$\Delta $ba}}=&\sin (\Delta a \lambda  t) \cos (\Delta b \lambda  t), & S_{\Delta b} S_{\Delta a}=&\sin (\Delta a \lambda t)\sin (\Delta b \lambda  t),\\
\chi_{AB}=&\lambda  t (\Delta a+\Delta b).
\end{align*}
This section includes the details of the simulations done for the dynamics of the two-qubit state when coupled to classical random fields.
\subsubsection{Application of the static noise}
 By using $Eq.\eqref{staticcommon}$, we get the final density matrix for common configuration model as:
\begin{equation}
\left[
\begin{array}{cccc}
 h_{11}^{ccm} & h_{12}^{ccm} & h_{12}^{ccm} & h_{11}^{ccm} \\
 h_{21}^{ccm} & h_{22}^{ccm} & h_{22}^{ccm} & h_{21}^{ccm} \\
 h_{21}^{ccm} & h_{22}^{ccm} & h_{22}^{ccm} & h_{21}^{ccm} \\
 h_{11}^{ccm} & h_{12}^{ccm} & h_{12}^{ccm} & h_{11}^{ccm}
\end{array}
\right]
\end{equation}
where
\begin{align*}
h_{11}^{ccm}=&\frac{4 t \text{$\Delta $m} \lambda +\text{Sin}[2 t (\text{$\Delta $m}-2 \text{$\Delta $o}) \lambda ]+\text{Sin}[2 t (\text{$\Delta $m}+2 \text{$\Delta $o}) \lambda ]}{16 t \text{$\Delta $m} \lambda },\\
h_{12}^{ccm}=&\frac{i (\text{Cos}[2 t (\text{$\Delta $m}-2 \text{$\Delta $o}) \lambda ]-\text{Cos}[2 t (\text{$\Delta $m}+2 \text{$\Delta $o}) \lambda ])}{16 t \text{$\Delta $m} \lambda },\\
h_{21}^{ccm}=&-\frac{i (\text{Cos}[2 t (\text{$\Delta $m}-2 \text{$\Delta $o}) \lambda ]-\text{Cos}[2 t (\text{$\Delta $m}+2 \text{$\Delta $o}) \lambda ])}{16 t \text{$\Delta $m} \lambda },\\
h_{22}^{ccm}=&\frac{4 t \text{$\Delta $m} \lambda +\text{Sin}\left[4 t \left(-\frac{\text{$\Delta $m}}{2}+\text{$\Delta $o}\right) \lambda \right]-\text{Sin}[2 t (\text{$\Delta $m}+2 \text{$\Delta $o}) \lambda ]}{16 t \text{$\Delta $m} \lambda }.\\
\end{align*}
Next, by using $Eq.\eqref{staticmixed}$, we get the final density matrix for the different configuration model as:
\begin{equation}
\left[
\begin{array}{cccc}
 p_{11}^{dcm} & p_{12}^{dcm} & p_{12}^{dcm} & p_{11}^{dcm} \\
 p_{21}^{dcm} & p_{22}^{dcm} &p_{22}^{dcm} & p_{21}^{dcm} \\
p_{21}^{dcm} & p_{22}^{dcm} & p_{22}^{dcm} & p_{21}^{dcm} \\
 p_{11}^{dcm} & p_{12}^{dcm} & p_{12}^{dcm}& p_{11}^{dcm}
\end{array}
\right]
\end{equation}
where
\begin{align*}
p_{11}^{dcm}=&\frac{-4 t^2 \text{$\Delta $m}^2 \lambda ^2+\text{Cos}[2 t (\text{$\Delta $m}-2 \text{$\Delta $o}) \lambda ]-2 \text{Cos}[4 t \text{$\Delta $o} \lambda ]+\text{Cos}[2 t (\text{$\Delta $m}+2 \text{$\Delta $o}) \lambda ]}{16 t^2 \text{$\Delta $m}^2 \lambda ^2},\\
p_{12}^{dcm}=&\frac{i (\text{Sin}[2 t (\text{$\Delta $m}-2 \text{$\Delta $o}) \lambda ]+2 \text{Sin}[4 t \text{$\Delta $o} \lambda ]-\text{Sin}[2 t (\text{$\Delta $m}+2 \text{$\Delta $o}) \lambda ])}{16 t^2 \text{$\Delta $m}^2 \lambda ^2},\\
p_{21}^{dcm}=&-\frac{i (\text{Sin}[2 t (\text{$\Delta $m}-2 \text{$\Delta $o}) \lambda ]+2 \text{Sin}[4 t \text{$\Delta $o} \lambda ]-\text{Sin}[2 t (\text{$\Delta $m}+2 \text{$\Delta $o}) \lambda ])}{16 t^2 \text{$\Delta $m}^2 \lambda ^2},\\
p_{22}^{dcm}=&\frac{4 t^2 \text{$\Delta $m}^2 \lambda ^2+\text{Cos}[2 t (\text{$\Delta $m}-2 \text{$\Delta $o}) \lambda ]-2 \text{Cos}[4 t \text{$\Delta $o} \lambda ]+\text{Cos}[2 t (\text{$\Delta $m}+2 \text{$\Delta $o}) \lambda ]}{16 t^2 \text{$\Delta $m}^2 \lambda ^2}.
\end{align*}
\subsubsection{Numerical results for the quantum correlations}
Here, we give numerical simulations for the dynamics of the decoherence, purity and concurrence given in $Eqs. \eqref{decoherence measure}$, $\eqref{purity measure}$ and $\eqref{concurrence}$ are given. These measures are further explored both for common and different configuration model from $Eqs. \eqref{staticcommon}$ and $\eqref{staticmixed}$ as:
\begin{align}
\mathcal{D}_{ccm}(t)=&\frac{1}{2} (\log[4]-\log[\frac{1}{2}-\frac{1}{4}\eta_1]-\log[2+\eta_1]-\frac{1}{\phi}{\arccoth [2\phi \text{csc}[2\phi]] \sin[2\phi]}),\\
\mathcal{D}_{dcm}(t)=& \frac{1}{2} (\log[2]-\log [\frac{1}{2}-\frac{1}{2}\eta_2]-\log[1+\eta_2]-\frac{1}{\phi^2}2 {\arctanh[\frac{\sin[\phi]^2}{\phi^2}] \sin[\phi]^2}),\\
\mathcal{P}_{ccm}(t)=&\frac{1}{16 \phi^2}{1+8 \phi^2-\cos[4\phi]},\\
\mathcal{P}_{dcm}(t)=&\frac{}{16 t^4 \phi^4}{3+8 \phi^4-4 \cos[2\phi]+\cos[4\phi]},\\
\mathcal{C}_{ccm}(t)=&\frac{1}{2 \sqrt{2}}{-\sqrt{4-\eta_3}+\sqrt{4+\eta_3}},\\
\mathcal{C}_{dcm}(t)=&-\frac{1}{\sqrt{2}}\sqrt{1-\eta_4}+\sqrt{1+\eta_4}.
\end{align}
where
\begin{align*}
\eta_1=&\frac{\sin[2\phi]^2}{\sqrt{\phi^2 \sin[2\phi]^2}}, &
\eta_2=&\frac{2 \sin[2\phi]^2}{\sqrt{\phi^2 \sin[2\phi]^2}},\\
\eta_3=&\frac{2 \sin[2 \phi]^2}{\sqrt{\phi^2 \sin[\phi]^2}},&
\eta_4=&\frac{\sin[\phi]^4}{\sqrt{\phi^4 \sin[\phi]^4}}\\
\phi=&t \Delta_m \lambda.
\end{align*}

\end{document}